\documentclass[
twocolumn,
prl,showpacs,preprintnumbers,amsmath,amssymb]{revtex4}

\usepackage{graphicx}
\usepackage{amssymb}
\usepackage{amsmath}
\usepackage{dsfont}
\usepackage{bm}

\newlength{\intwidth}
\DeclareRobustCommand{\fpint}[2]
  {\mathop{%
     \text{%
       \settowidth{\intwidth}{$\int$}%
       \makebox[0pt][l]{\makebox[\intwidth]{$-$}}%
       $\int_{#1}^{#2}$}}}

\begin{document}

\title{Slow-light solitons}

\author{Ulf Leonhardt}

\affiliation{School of Physics and Astronomy, University of St
  Andrews, North Haugh, St Andrews, KY16 9SS, Scotland }     

\begin{abstract}
A new type of soliton with controllable speed is constructed
generalizing the theory of slow-light propagation 
to an integrable regime of nonlinear dynamics.
The scheme would allow the quantum-information transfer
between optical solitons and atomic media. 
\end{abstract}


\maketitle

\begin{figure}[t]
\begin{center}
\includegraphics[width=20.0pc]{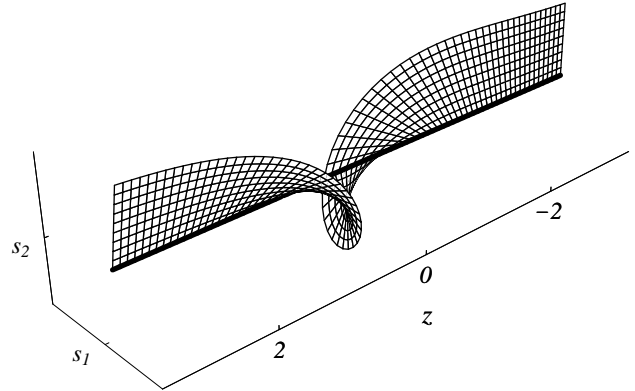}
 \caption{
\small
A slow-light soliton is a polarization twist.
To visualize the polarization state of light
we use the Stokes parameters \cite{BornWolf}
$s_1$, $s_2$ and $s_3$
(with $s_3$ in $z$ direction)
defining a vector at each point in space $z$ and time $t$.
For a given time $t$ 
we plot a strip spanned by $z$ and the polarization vector
in arbitrary units.
When time progresses, 
the polarization twist moves through the medium 
with velocity (\ref{eq:velocity}).
The figure illustrates the spatial polarization 
profile at time $t=0$ 
in spatial units of the soliton length (\ref{eq:length}).
The Stokes parameters \cite{BornWolf} are calculated 
from the solution (\ref{eq:results})
for a slow-light soliton
with linear initial polarization,
Rabi frequency
$\Omega=0.5 {\rm MHz}$,
spectral parameter
$\xi+i\eta=10.0\times\exp(i0.4\pi) \,{\rm MHz}$,
$Q_0=\Phi_0=0$
and for a sharp line with $\nu(\Delta)=\delta(\Delta)$.
} 
\label{fig:twist}
\end{center}
\end{figure}

Solitons \cite{Russell,Steudel,Bullough,Ablowitz}
are stable wave packets occurring in many areas 
of the physical world, from tsunamis,
flocks of migrating arctic geese \cite{Abram}
to light pulses in optical fibres \cite{Abram,Leuchs}.
Their stability stems from the non-linear balance 
between dispersing and focusing processes.
As a feature of non-linear dynamics,
the speed  of a soliton may depend on its amplitude.
For example, the height of a solitary water wave
determines its velocity \cite{Russell}.
However, once a soliton is launched there is usually no
further control over its speed.
Here we show how to generate optical solitons in
atomic media that can be slowed down or accelerated at will.
Ultimately, this method will allow the storage, retrieval
and possibly the manipulation of the quantum information 
\cite{Braunstein} of solitons
\cite{Abram,Leuchs,Drummond,LaiHaus} 
in media. 
This idea extends the simplest scheme \cite{Liu,Philips}
for slow-light propagation 
\cite{Hau,Slow} 
to a genuinely non-linear regime.\\
\indent
In the pioneering experimental demonstrations
\cite{Liu,Philips,Hau} of slow light 
\cite{Slow} 
atomic media store the shape of light pulses in the
spin states of atoms \cite{Fleisch}
in what is known as dark states \cite{Arimondo}.
Usually, such experiments operate in a regime
where the atomic spins deviate only slightly 
from a default direction,
a regime of linear spin waves \cite{Fleisch}.
Here we consider strong spin modulations
where, as we show, the non-linear dynamics
of light and atoms creates polarization solitons,
see Fig.\ \ref{fig:twist} for an illustration.
Unlike other solitons known so far,
their speed can be controlled 
after they have been launched,
in precisely the same way as for slow light.
In quantum-information applications \cite{Braunstein},
such slow-light solitons 
could complement the use of quantum solitons in fibres 
\cite{Abram,Leuchs}
with the advantage of storing quantum information in media 
and complement methods for quantum memory
\cite{Qmem}
with the advantages of non-linear dynamics,
in particular the intrinsic stability of solitons.
Strong spin polarizations can be imaged  
by illuminating the sample from the side 
with resonant light of uniform polarization.
Moreover, 
they could be manipulated with light or magnetic fields, 
possibly a step from quantum-information storage
to quantum-information processing in atomic vapours.

Consider a cell filled with atomic vapour \cite{Philips} or a 
Bose-Einstein condensate \cite{Liu}
illuminated with light propagating in one direction.
Only three atomic levels shall interact with the light,
an excited state and two degenerate ground states,
for example two hyperfine levels \cite{Liu,Philips}.
As a consequence of the conservation of angular momentum,
the two circular polarizations of light couple to either one
of the ground states.
This type of interaction between three-level atoms and light is
called $\Lambda$ configuration \cite{Maimistov}.
In the presence of light of constant but arbitrary polarization
the atoms relax by spontaneous emission to spin states
pointing in the opposite direction of the optical polarization, 
so-called dark states \cite{Arimondo},
a process known as optical pumping.
The light spin-polarizes the atoms. 
Once the atoms are in dark states, they follow the
polarization of light by themselves, 
without spontaneous emission,
as long as their excited states are sparsely populated
\cite{Leo}.
Suppose that the polarization of the light has varied
in time and has imprinted a spatial modulation in the
atomic spin distribution.
Any subsequent light incident with uniform polarization will
attempt to re-polarize the atoms, but, in turn, 
is re-polarized itself.
The tail-end of the spin profile is reset to the default 
polarization and the front-end moves.
The light thus carries the spin modulation along the medium,
possibly modulating it further,
with a speed that depends on the intensity.
When the light is switched off,
the spin profile stands still, as long as it is not eroded by
diffusion, de-polarizing atomic collisions or the spin
precession in stray magnetic fields.
Illuminating the atoms will continue to move the spin wave
until it reaches the boundary of the medium where 
the stored spin profile emerges as a polarization profile of light.
The pioneering demonstrations \cite{Liu,Philips} 
of stopping light \cite{Slow} 
have used this method \cite{Fleisch},
but with an important qualification \cite{Fleisch}:
One of the polarization modes of light dominates the other. 
In this regime, the spin-perturbation waves are small, 
with linear dynamics, 
and they move without changing shape \cite{Fleisch}.
In the case of strong polarization modulations,
the balance between the two competing processes involved,
light polarizing the atoms and 
the atoms polarizing the light,
may generate stable polarization structures, 
solitons \cite{Steudel,Bullough,Ablowitz}.

To develop a quantitative description of slow-light solitons,
we assume that the light propagates in positive
$z$ direction with amplitudes that vary slowly in comparison
with the carrier frequency $\omega$. 
We describe the left and right circular polarization amplitudes
in terms of the Rabi frequencies \cite{Allen} 
$\Omega_+(t,z)$ and $\Omega_-(t,z)$, respectively, {\it i.e.}\
by the transversal components of the electric field strength
divided by $\hbar$ and
multiplied by the dipole moment $\kappa$ of the
atomic transitions the light interacts with.
We distinguish atoms with different detunings $\Delta$
of the transition frequencies from resonance with the light,
caused, for example, 
by the Doppler effect of their thermal motion.
We describe the quantum state of each atom in terms of
the density matrices $\rho(t,z,\Delta)$
considering only the ground states $|\pm\rangle$
and the excited state $|e\rangle$ that interact with the light.
In the slowly varying envelope approximation \cite{Allen}
the light amplitudes obey the 
approximative Maxwell equations 
\begin{equation}
\label{eq:maxwell}
\frac{\partial\Omega_\pm}{\partial t} +
c\frac{\partial\Omega_\pm}{\partial z}=
ig \int_{-\infty}^{+\infty}
\langle e |\, \rho\, | \pm\rangle\,\nu\,d\Delta
\,,\,\,
g  = 
\frac{\kappa^2\omega}{4\hbar\varepsilon_0}\,n
\,.
\end{equation}
Here $n(z)$ denotes 
the atom-number density profile of the medium
and $\nu(\Delta)$ the relative distribution
of the detuned atoms.
$\varepsilon_0$ is the electric permeability of the vacuum.
At this stage we neglect any atomic relaxation,
but we determine later the condition
when this is justified. 
Without relaxation the atomic density matrices 
evolve according 
to the Liouville equation
in the interaction picture
\begin{eqnarray}
\label{eq:liouville}
\frac{\partial\rho}{\partial t}
&=& -i[H,\rho]
\,,\nonumber\\
H &=&  -\Delta\, |e\rangle \langle e|
-\sum_\pm\bigg(
\frac{\Omega_\pm^*}{2} |\pm\rangle \langle e| + 
\frac{\Omega_\pm}{2} |e\rangle \langle \pm|
\bigg)
\,.
\end{eqnarray}
Despite the apparent complexity and non-linearity of the
Maxwell-Liouville equations 
(\ref{eq:maxwell},\ref{eq:liouville}),
they belong to the class of integrable systems \cite{Maimistov}.
The key element of an integrable system is a Lax pair
\cite{Steudel,Bullough,Ablowitz}
of matrices $U$ and $V$
that generates the equations of motion as the
compatibility condition 
$\partial U/\partial\zeta - \partial V/\partial\tau +[U,V] = 0$
for a variable complex spectral parameter,
which allows the use of the Inverse Scattering Method
\cite{Steudel,Bullough,Ablowitz}.
For the three-level Maxwell-Liouville system 
(\ref{eq:maxwell},\ref{eq:liouville})
the detuning $\Delta$ serves as the spectral parameter,
$\tau$ means the retarded time $t-z/c$,
and $\zeta$ denotes the spatial coordinate $z/c$.
One verifies that the Lax pair is
\cite{Lax}
\begin{equation}
\label{eq:lax}
U = -iH \,,\quad
V=-\frac{ig}{2}\fpint{-\infty}{+\infty}
\frac{\rho(\Delta')\nu(\Delta')\, d\Delta'}{\Delta-\Delta'} 
\,.
\end{equation}
Therefore, the theory of slow light is integrable
\cite{Steudel,Bullough,Ablowitz},
which comes in useful for finding analytic solutions,
and, more importantly, which allows the existence of
stable non-linear waves, solitons.

Various methods for finding soliton solutions 
start from a Lax pair of type (\ref{eq:lax}).
We use a modification of the
Zakharov-Shabat dressing method \cite{ZSh}
for a variable background field  
$\Omega(\tau)$ of constant polarization
with the atoms being in the
corresponding dark states.
We obtain 
the single-soliton solution for the field 
$\Omega_\pm$ 
and the atoms $\rho=\psi_a\psi^*_b$ as
\begin{eqnarray}
\Omega_\pm &=& \Omega \varphi_\pm 
\,,\nonumber\\
\varphi_+ &=& \frac{\xi-i\eta\tanh Q}{\xi+i\eta}
\,,\quad
\varphi_- = \frac{\eta\,\exp(-i\Phi)\,{\rm sech}\, Q}{\xi+i\eta}
\,,\quad
\nonumber\\
Q &=& Q_0 - \int
\frac{\eta\,|\Omega|^2 d\tau}{4(\xi^2+\eta^2)}
+\int\int
\frac{\eta g \nu\,d\Delta\, d\zeta}{2((\xi-\Delta)^2+\eta^2)}
\,,\quad
\nonumber\\
\Phi &=& \Phi_0 + \int
\frac{\xi\,|\Omega|^2 d\tau}{4(\xi^2+\eta^2)}
-\int\int 
\frac{(\xi-\Delta) g \nu\,d\Delta\, d\zeta}{2((\xi-\Delta)^2+\eta^2)}\,,
\nonumber\\
\psi_+ &=& -\frac{(\xi+i\eta)\varphi_-}{\xi-\Delta+i\eta}
\,,\quad
\psi_- = \frac{(\xi+i\eta)\varphi_+-\Delta}{\xi-\Delta+i\eta}
\,,\nonumber\\
\psi_e &=& \frac{\Omega_-}{2(\xi-\Delta+i\eta)}
\label{eq:results}
\end{eqnarray}
in the limit $\xi^2+\eta^2\gg |\Omega|^2$ where
$\xi$, $\eta$, $Q_0$, $\Phi_0$ are real constants 
and $\Omega(\tau)$ is a complex function.
One verifies that the results (\ref{eq:results}) 
solve the reduced Maxwell equation (\ref{eq:maxwell}) 
and the approximate equation  (A13) of Ref.\  \cite{Leo}
that describes the atomic dynamics.
In the solution (\ref{eq:results}) the incident field is left-circularly
polarized. We can, however, apply the transformation 
\begin{equation}
\label{eq:poltrans}
\left(
    \begin{array}{c}
      \Omega_+' \\
      \Omega_-'
    \end{array}
\right)
=
B
\left(
    \begin{array}{c}
      \Omega_+\\
      \Omega_-
    \end{array}
\right)
\,,\,\,
\left(
    \begin{array}{c}
      \psi_+' \\
      \psi_-'
    \end{array}
\right)
=
B^\top
\left(
    \begin{array}{c}
      \psi_+\\
      \psi_-
    \end{array}
\right)
\end{equation}
with constant unitary matrices $B$
to describe an arbitrary initial polarization state.
The results (\ref{eq:results}) are generalizations 
of some known solutions \cite{ParkByrne} to 
slow-soliton propagation in the presence of
inhomogeneous line broadening $\nu(\Delta)$,
variable spatial density $n(\zeta)$ and,
more importantly, time-dependent background 
$\Omega(\tau)$.

Slow-light solitons turn out to be pure polarization 
structures that, 
above a critical strength, perform complete 
polarization twists, see Fig.\ \ref{fig:twist}.
Let us discuss their properties 
based on the analytic solution (\ref{eq:results}).
A single soliton depends on four real parameters
$\xi$, $\eta$, $Q_0$, $\Phi_0$ and on one arbitrary complex
function $\Omega(\tau)$ for which we require
that $|\Omega|^2 \ll |\xi|^2+|\eta|^2$.  
We obtain from the solution (\ref{eq:results}) that
\begin{equation}
|\Omega_+(t,z)|^2+ |\Omega_-(t,z)|^2=|\Omega(t-z/c)|^2 \,,
\end{equation}
which indicates that $|\Omega|^2$ describes the total intensity
and that any incident intensity modulation
propagates through the medium at the speed of light in vacuum.
The parameter $Q_0$
characterizes the position and  $\Phi_0$ the phase
of the polarization deformation,
whereas $\xi$ and $\eta$ constitute
the complex spectral parameter $\xi + i\eta$
whose argument $\theta$ describes the maximal
polarization deviation.
Polarization twists occur when $|\theta|\ge\pi/2$.
Beyond the soliton the light returns to the
incident polarization state, acquiring a geometric phase 
\cite{Berry} of $2\theta$.
The atoms are in pure quantum states 
with probability amplitudes $\psi_a(t,z,\Delta)$.
In regions of nearly uniform optical polarization
the atoms are essentially in dark states,
but their ability to follow
a changing polarization of light is limited due to the 
detuning $\Delta$.
Outside the medium the solution (\ref{eq:results}) 
for $g=0$ describes the shape
of the polarization pulse required to launch the soliton.
In the medium, the soliton propagates with velocity
\begin{equation}
\label{eq:velocity}
\frac{v}{c}=\frac{|\Omega|^2}{2(\xi^2+\eta^2)}
\left(\int_{-\infty}^{+\infty} 
\frac{g\nu(\Delta)\,d\Delta}{(\xi-\Delta)^2+\eta^2}\right)^{-1}
\sim\frac{|\Omega|^2}{2g}
\end{equation}
for $v/c\ll 1$ and in the second approximate expression 
for $\xi^2+\eta^2$ much larger than 
a characteristic detuning $\Delta^2$.
We obtain from the solution (\ref{eq:results}) the 
length scale of the polarization profile, the soliton length
\begin{equation}
\label{eq:length}
l_s=\frac{4(\xi^2+\eta^2)}{\eta\,|\Omega|^2}\,v
\sim
\frac{2c(\xi^2+\eta^2)}{\eta g}
\,.
\end{equation}
Typically for solitons \cite{Steudel,Bullough,Ablowitz},
the length is adjusted to the spectral parameter. 
Atypically \cite{Steudel,Bullough,Ablowitz}, 
the speed hardly depends on any
parameters, except for the external light intensity
in exactly the same way 
as for traditional slow light \cite{Slow},
allowing control over the soliton after it has been launched.
  
A critical issue for slow-light solitons is the potential threat of
absorption, in contrast to fast self-induced-transparency solitons
\cite{Steudel,Maimistov} that excite and de-excite 
the atoms on times much shorter than
the relaxation-time scales of the atoms.
Consider a slow-light soliton of length $l_s$
traversing the distance $l$ in the medium.
The principal relaxation mechanism is spontaneous emission 
from the excited state $|e\rangle$.  
The population of $|e\rangle$ is largest for zero detuning where 
$|\psi_e|^2  \sim 2(c/g)l_s^{-2} v$
at the peak of the soliton. 
We use this case to estimate the absorption.
First, we express $g$ 
defined in the Maxwell equations (\ref{eq:maxwell})
in terms of the Einstein $A$ coefficient \cite{Loudon}.
We get $g = {3}/({16\pi})\,cA\, n\lambda^2$,
where $\lambda$ denotes the optical wavelength
$\lambda = 2\pi c/ \omega$.
In practice \cite{Liu,Philips}
the excited state decays into three ground-state
levels, not only the $|\pm\rangle$ selected by 
the propagation direction of light.
We estimate the fractional loss $\eta_L$ 
as $3A$ multiplied
by the propagation time $l/v$ times
the maximal excited-state population, and obtain
\begin{equation}
\eta_L = 
\frac{32\pi}{n\lambda^3}\, 
\frac{l\lambda}{l_s^2}
\,.
\end{equation}
In the experiments \cite{Liu,Philips}
$n\lambda^3$ has been in the order of $1$ or larger.
Losses are thus negligible for solitons much longer
than the geometric mean of wavelength and distance travelled,
leaving enough room for soliton propagation.
This requirement resembles the condition for traditional
slow light \cite{Slow} where the pulses are restricted in 
frequency space, which amounts to a minimal length 
scale in ordinary space. 

Having established that slow-light solitons are feasible,
we identify their quantum properties.
Fundamentally, the solitons are quantum excitations
of the electromagnetic field interacting with the atoms
of the medium, but they are very close to the classical 
limit (\ref{eq:results}).
Their quantum nature manifests itself in fluctuations
around the classical amplitudes.
A classical soliton is stable against fluctuations 
except when the fluctuations occur 
in its degrees of freedom, its parameters,
in our case in any $\alpha$ $\in$ $\{\xi,\eta,Q_0,\Phi_0\}$,
because such parameter fluctuations transform 
the soliton into another stable soliton.
The Goldstone modes generated by infinitesimal 
parameter changes define the relevant quantum modes 
of the soliton \cite{LaiHaus}.
Therefore, we represent the quantum fields 
$\hat{\Omega}_\pm$ and $\hat{\psi}_a$ 
of light and atoms as 
\begin{eqnarray}
\hat{\Omega}_\pm &=& \Omega_\pm + \sum_\alpha 
\frac{\partial\Omega_\pm}{\partial\alpha}\,
(\hat{\alpha}-\alpha)
\,,\nonumber\\
\hat{\psi}_a &=& \psi_a + \sum_\alpha 
\frac{\partial\psi_a}{\partial\alpha}\,
(\hat{\alpha}-\alpha)
\,.
\label{eq:model}
\end{eqnarray}
The quantum properties of the soliton modes 
depend on the parameterization of the classical soliton,
chosen here (\ref{eq:results}) such that they are 
particularly simple.
We deduce the commutation relations of the mode operators
$\hat{\alpha}$ from the quantum field theory of light \cite{Loudon}
in the slowly-varying envelope approximation \cite{Allen}.

Quantum field theories
may start from a fundamental quantum commutator
and a classical mode decomposition with an appropriate
scalar product.
Here we describe the field by two components 
$\hat{A}_\pm$ of the vector potential 
in the Coulomb gauge \cite{Loudon} and SI units
with the equal-time commutator \cite{Loudon}
$[\hat{A}_\pm(t,z_1),
{\varepsilon_0\sigma}\,
{\partial \hat{A}_\pm(t,z_2)}/{\partial t}]  =
i\hbar\, \delta(z_1-z_2)$
where $\sigma$ denotes the longitudinal cross section
of the medium. 
In the slowly-varying envelope approximation \cite{Allen}
the Rabi frequencies $\Omega_\pm$ are, 
apart from a prefactor $\kappa/\hbar$, 
the components of the electric field strength
$-\partial A_\pm/\partial t$ that propagate 
with the carrier $\exp(-i\omega\tau)$.
Therefore we define 
$\hat{\Omega}_\pm =$
$\kappa/(2\hbar)\,\exp(-i\omega\tau)$
$(i\omega\hat{A}_\pm-\partial \hat{A}_\pm/\partial t)$
and get the commutation relation 
in retarded time
\begin{equation}
\label{eq:commutator}
[\hat{\Omega}_\pm(\tau_1,\zeta), \,
\hat{\Omega}_\pm^\dagger(\tau_2,\zeta)]  =
\frac{\kappa^2\omega}
{2\varepsilon_0\hbar\,\sigma c}\, 
\delta(\tau_1-\tau_2)
\,.
\end{equation}
Inspired by Lai's and Haus' quantum theory of 
self-induced-transparency solitons  \cite{LaiHaus}
we define the scalar product
of the fluctuation modes
in the style of Poisson brackets
\begin{equation}
\{\alpha,\beta\} \equiv \frac{1}{8i}
\int_{-\infty}^{+\infty} \sum_\pm \left(
\frac{\partial \Omega_\pm^*}{\partial \alpha} 
\frac{\partial \Omega_\pm}{\partial \beta} -
\frac{\partial \Omega_\pm}{\partial \alpha} 
\frac{\partial \Omega_\pm^*}{\partial \beta}
\right) d\tau
\,.
\end{equation}
One verifies by straightforward but lengthy calculations
that the solutions (\ref{eq:results}) are designed
such that all $\{\alpha,\beta\}$ vanish, except 
$\{Q_0,\xi\}  =  -\{\xi,Q_0\} = \{\Phi_0,\eta\} = -\{\eta,\Phi_0\} = 1$.
These orthonormality conditions serve to deduce 
the commutation relations of the fluctuation operators
from the commutator (\ref{eq:commutator}).
We re-scale the quantum spectral parameter as
$\hat{\xi}+i\hat{\eta} = 
(\hat{\xi}_0+i\hat{\eta}_0)\,
\kappa^2\omega/(16\varepsilon_0\hbar\,\sigma c)$.
Following the procedure by Lai and Haus \cite{LaiHaus}
we obtain 
the Heisenberg commutation relations 
\begin{eqnarray}
i &=&
[\hat{Q}_0, \hat{\xi}_0] = [\hat{\Phi}_0, \hat{\eta}_0] \,
\nonumber\\
0 &=&
[\hat{\xi}_0, \hat{\eta}_0] = 
[\hat{Q}_0, \hat{\Phi}_0] =
[\hat{Q}_0, \hat{\eta}_0] =
[\hat{\Phi}_0, \hat{\xi}_0] 
\,.
\label{eq:heisenberg}
\end{eqnarray}
The components of the quantum spectral parameter, 
$\hat{\xi}_0$ and $\hat{\eta}_0$,
thus represent the canonical momenta of two independent
Heisenberg pairs with $\hat{Q}_0$ and $\hat{\Phi}_0$
as their partners.
This implies that the position $Q_0$ of the soliton
is complementary to $\xi_0$ and the phase $\Phi_0$
of the polarization twist is complementary to 
its magnitude $\eta_0$.
The quantum properties of these two modes
encode the quantum information \cite{Braunstein}
of the soliton, 
regardless of whether it is moving or frozen.
The overall amplitude $\Omega$
plays the role of the 
control beam \cite{Liu,Philips}. 
When the soliton is stopped 
by switching off the incident light
the quantum information is stored in the medium,
since the atomic modes
$\partial\psi_a/\partial\alpha$
do not vanish when $\Omega(\tau) \rightarrow 0$.
Switching on the light will set the soliton in motion
and will release the stored quantum information 
in the polarization of the light emerging 
at the boundary of the medium.

{\it Summary. ---}
The theory of slow light \cite{Liu,Philips} 
is integrable \cite{Steudel,Bullough,Ablowitz}.
A slow-light soliton is a polarization structure
propagating with a speed that is proportional to
the total intensity of the incident light.
Entire solitons can be stopped and retrieved
in atomic media with their quantum information
stored in atomic coherences.

I am grateful to Ildar Gabitov, Landau Institute,
for teaching me soliton theory
and to Ilya Vadeiko, St Andrews,
for checking the calculations of this paper.
The research was supported by 
the Leverhulme Trust, 
EPSRC, 
and the ESF Programme Cosmology in the Laboratory.

\end{document}